\documentclass[12pt]{iopart}

\newread\testifexists
\def\GetIfExists #1 {\immediate\openin\testifexists=#1
    \ifeof\testifexists\immediate\closein\testifexists\else
    \immediate\closein\testifexists\input #1\fi}

\usepackage{gthstyle}\usepackage{amsfonts}
\usepackage{amssymb}
\immediate\openin\testifexists=e
    \ifeof\testifexists\immediate\closein\testifexists\else
    \immediate\closein\testifexists\input e\fipsf

\def\Bbb#1{\setbox0=\hbox{$\tt #1$}  \copy0\kern-\wd0\kern .1em\copy0}

\def\bbf#1{\setbox0=\hbox{$#1$} \kern-.025em\copy0\kern-\wd0
        \kern.05em\copy0\kern-\wd0 \kern-.025em\raise.0433em\box0}

\def\a{\alpha}      \def\b{\beta}         
\def\d{\delta}      \def\D{\Delta}  
               \def\L{\Lambda}
\def\m{\mu}                     
         \def\j{\psi}    
\def\r{\varrho}

\def\w{\omega}        

  \def\OO{{\cal O}}
\def\pa{\partial} \def\ra{\rightarrow}

\def\dd{{\rm d}}  \def\bra{\langle}   \def\ket{\rangle}

\def\deff{\ {\buildrel{\rm def}\over{=}}\ }

\def\fract#1#2{{\textstyle{#1\over#2}}}
\def\ffract#1#2{\raise .3 em\hbox{$\scriptstyle#1$}\kern-.25em/
                \kern-.2em\lower .2 em \hbox{$\scriptstyle#2$}}

\def\half{\fract12}  

\def\part#1#2{{\partial#1\over\partial#2}}

\def\iss{\ =\ }

\newcommand{\be}{\begin{eqnarray}}
\newcommand{\ee}{\end{eqnarray}}
\newcommand{\eqn}[1]{(\ref{#1})}

\newcommand{\bitm}[1]{\begin{itemize}\item[#1]} 
\newcommand{\ei}{\end{itemize}}

\newcommand{\fn}{\footnote}

\renewcommand{\thesection}{\arabic{section}}
\renewcommand{\theequation}{\thesection.\arabic{equation}}
\newcommand{\newsec}[1]{\section{#1}\setcounter{equation}{0}}

\newcommand{\eel}[1]{\label{#1}\end{eqnarray}}\newcommand{\crl}[1]{\label{#1}\\ }  

\begin{document}

\title{The mathematical basis for deterministic quantum mechanics} \author{Gerard 't~Hooft}
\address{Institute for Theoretical Physics \\
Utrecht University, Leuvenlaan 4\\ 3584 CC Utrecht, the Netherlands\medskip \\ and \medskip \\
Spinoza Institute \\ Postbox 80.195 \\ 3508 TD Utrecht, the Netherlands \smallskip}
\eads{\mailto{g.thooft@phys.uu.nl}}

\begin{quotation} \noindent {\large\bf Abstract} \medskip \\
If there exists a classical, \emph{i.e.} deterministic theory underlying quantum mechanics, an explanation must be
found of the fact that the Hamiltonian, which is defined to be the operator that generates evolution in time, is
bounded from below. The mechanism that can produce exactly such a constraint is identified in this paper. It is the
fact that not all classical data are registered in the quantum description. Large sets of values of these data are
assumed to be indistinguishable, forming \emph{equivalence classes}. It is argued that this should be attributed to
information loss, such as what one might suspect to happen during the formation and annihilation of virtual black
holes.

The nature of the equivalence classes is further elucidated, as it follows from the positivity of the Hamiltonian. Our
world is assumed to consist of a very large number of subsystems that may be regarded as approximately independent, or
weakly interacting with one another. As long as two (or more) sectors of our world are treated as being independent,
they all must be demanded to be restricted to positive energy states only. What follows from these considerations is a
unique definition of energy in the quantum system in terms of the periodicity of the limit cycles of the deterministic
model.
\end{quotation}

\newsec{Introduction}\label{intro.sec} There may exist different versions of deterministic theories
underlying Quantum Mechanics, usually referred to as ``hidden variable theories". For instance, one
may or may not assume the occurrence of information loss at tiny distance scales. One may suspect
some sort of cellular automaton or a classical system of continuous fields, or even classical
loops, \(D\)-branes, or whatever. Instead of quantizing such systems in the usual manner, we here
consider what we will refer to as \emph{pre-quantization}. With this term we mean that the physical
system is not modified, as in usual quantization schemes, which depend on a new constant \(\hbar\),
but only rephrased in a language suitable for quantum mechanical manipulations at a later stage.

Pre-quantization may be useful when complex systems are handled probabilistically. The probability distribution \(W\)
is then re-written as the absolute square of a wave function. The wave function obeys all the axioms of conventional
quantum mechanics, allowing us to perform all the mathematical tricks known from Quantum Mechanics and Quantum Field
Theory, such as group representation theory and renormalization group transformations.

We suspect that our world can be understood by starting from a pre-quantized classical, or `ontological', system. However, a
serious difficulty is then encountered: one indeed gets Quantum Mechanics, but the Hamiltonian is not naturally bounded from
below. If time would be assumed to be discrete, the Hamiltonian eigenvalues would turn out to be periodic, so one might limit
oneself to eigenvalues \(E\) with \(0\le E<2\pi/\d t\), where \(\d t\) is the duration of a fundamental time step, but then the
choice of a vacuum state is completely ambiguous, unlike the situation in the real world that one might want to mimic. If time is
continuous, the Hamiltonian eigenvalues tend to spread over the real line, from \(-\infty\) to \(\infty\).

In realistic theories, one therefore must impose a ``superselection rule", projecting out a subspace of Hilbert space
where all energies are non-negative. How exactly to do this will be described here. At first sight, the freedom to
choose phase factors in wave functions allows one to make such a selection without loss of generality. This
observation, however, is not the solution to the positivity problem of the Hamiltonian, since positivity must also be
demanded to hold for subsystems, and when such systems interact, the suppression of negative energy states might cause
the violation of unitarity, or locality, or both.

In this paper, we derive the plausibility of our assumptions from first principles. First, the formalism is displayed
in Section~\ref{beables.sec}. Deterministic systems are shown to be accessible by quantum mechanical procedures,
although this does not turn them into acceptable quantum mechanical models just yet, because the Hamiltonian is not
bounded from below. Then, we demonstrate that the most basic building blocks of any deterministic theory consists of
units that would evolve with periodicity if there were no interaction (Section~\ref{harmosc.sec}). We use the
empirically known fact that the Hamiltonians are all bounded from below both before introducing the interaction and
after having included the interaction. This necessitates our introduction of equivalence classes
(Section~\ref{twoormore.sec}), such that neither the quantum mechanical nor the macroscopic observer can distinguish
the elements within one equivalence class, but they can distinguish the equivalence classes.

This procedure is necessary in particular when two systems are considered together \emph{prior} to considering any
interaction. We are led to the discovery that, besides the Hamiltonian, there must be a classical quantity \(E\) that
also corresponds to energy, and is absolutely conserved as well as positive (Section~\ref{hamilton.sec}). It allows us
to define the equivalence classes. We end up discovering a precise definition of the quantum wave function for a
classical system (both amplitude and phase), and continue our procedure from there.

Physical and intuitive arguments were displayed in Ref.~\cite{disdet}. In that paper, it was argued that any system
with information loss tends to show periodicity at small scales, and quantization of orbits. It was also argued that
some lock-in mechanism was needed to relate the Hamiltonian with an ontologically observable quantity \(E\) that is
bounded from below. The lock-in mechanism was still not understood; here however we present the exact mathematical
treatment and its relation to information loss. Interaction can be introduced in a rather direct manner (Section
\ref{interaction.sec}), by assuming energy not to be directly additive, but then it is difficult to understand how
different energy sectors of the theory can be related to one another.

A more satisfactory picture emerges if one realizes that energy is not directly locally observable, but determined by
the periods of the limit cycles. This is explained in Section~\ref{cycles.sec}. We think that this interpretation is
imperative, and it sheds an interesting new light on the phenomenon we call quantum mechanics. After a discussion of
our results (Section~\ref{disc.sec}) an appendix follows in which we discuss the `random automaton'. It allows us to
estimate the distribution of its limit cycles, though we immediately observe that the quantum models it generates are
not realistic because the energy will not be an extensive quantity. The deterministic models that might reproduce
observed quantum field systems must be very special.

This paper was written while these facts were being discovered, so that it represents an original train of thought,
which may actually be useful for the reader.

\newsec{Variables, beables and changeables}\label{beables.sec} Any classical, deterministic system will contain some
set of degrees of freedom \(\vec q\) that follow some orbit \(\vec q(t)\) in time. Time might be
defined as a discrete variable or a continuous one, but this distinction is not as fundamental as
one might think. If time is discrete, then the set \(\vec q\) will have to include a clock that
gives a tick at every time step \(t_n=n\,\d t\), or, \be {\dd\over\dd t}\ q_\mathrm{clock}&=&1\ ;\\
q_i\ra q_i'(\vec q)\ &\mathrm{at}\ &q_\mathrm{clock}=0\ \mathrm{mod}\ \d t\ ,\quad\forall i\ne \mathrm{clock}\
.\eel{clock} It is not difficult to ascertain that this is just a special case of a more general equation of motion,
\be {\dd\over\dd t}\ \vec q=\vec f(\vec q)\ .\eel{eom} For simplicity we therefore omit specific references to any
clock\fn{Thus, we do, as yet, use an absolute notion of time. Special and general relativistic transformations are left
for future studies.}. In general, the orbit \(\vec q(t)\) will be dictated by an equation of motion of the form
\eqn{eom}.

In the absence of information loss, this will correspond to a Hamiltonian \be H=\sum_ip_if_i(\vec q)+g(\vec q)\ ,
\eel{hamil} where \(p_i=-i\pa/\pa q_i\) is the \emph{quantum} momentum operator. It will be clear that the quantum
equations of motion generated by this Hamiltonian will exactly correspond to the \emph{classical} equation \eqn{eom}.
The function \(g(\vec q)\) is arbitrary, its imaginary part being adjusted so as to ensure hermiticity: \be
H-H^\dagger\iss -i\vec\nabla\cdot\vec f(\vec q)+2\,i\,\mathrm{Im}(g(\vec q))\iss 0\ , \eel{hermitean}

Any observable quantity \(A(\vec q)\), not depending on operators such as \(p_i\), and therefore
commuting with all \(q_i\), will be called a \emph{beable}. Through the time dependence of \(\vec
q\), the beables will depend on time as well. Any pair of beables, \(A\) and \(B\), will commute
with one another at all times: \be[A(t_1),\,B(t_2)]=0\ ,\quad\forall\, t_1,\,t_2\ . \eel{beables}

A \emph{changeable} is an operator not commuting with at least one of the \(q_i\)'s. Thus, the operators \(p_i\) and
the Hamiltonian \(H\) are changeables. Using beables and changeables as operators\cite{disdet}, we can employ all
standard rules of quantum mechanics to describe the classical system \eqn{eom}. At this point, one is tempted to
conclude that the classical systems form just a very special subset of all quantum mechanical systems.

This, however, is not quite true. Quantum mechanical systems normally have a Hamiltonian that is bounded from below;
the Hamiltonian \eqn{hamil} is not. At first sight, one might argue that all we have to do is \emph{project out} all
negative energy states\cite{disdet}\cite{Elze1}. We might obtain a physically more interesting Hilbert space this way,
but, in general, the commutator property \eqn{beables} between two beables is lost, if only positive energy states are
used as intermediate states. As we will see, most of the beables \eqn{beables} will not be observable in the quantum
mechanical sense, a feature that they share with non-gauge-invariant operators in more conventional quantum systems
with Yang-Mills fields. The projection mechanism that we need will be more delicate. As we will see, only the beables
describing equivalence classes will survive as quantum observables.

We will start with the Hamiltonian \eqn{hamil}, and only later project out states. Before projecting out states, we may
observe that many of the standard manipulations of quantum mechanics are possible. For instance, one can introduce an
integrable approximation \(f^{(0)}_i(\vec q)\) for the functions \(f_i(\vec q)\), and write \be f_i(\vec
q)=f^{(0)}_i(\vec q)+\d f_i(\vec q)\ , \eel{pertf} after which we do perturbation expansion with respect to the small
correction terms \(\d f_i\). However, the \emph{variation principle} in general does not work at this level, because it
requires a lowest energy state, which we do not have.

\newsec{The harmonic oscillator}\label{harmosc.sec}

We assume that a theory describing our world starts with postulating the existence of sub-systems
that in some first approximation evolve independently, and then are assumed to interact. For
instance, one can think of independent local degrees of freedom that are affected only by their
immediate neighbors, not by what happens at a distance, baring in mind that one may have to expand
the notion of immediate neighbors to include variables that are spatially separated by distances of
the order of the Planck length. Alternatively, one may think of elementary particles that, in a
first approximation, behave as free particles, and are then assumed to interact.

Temporarily, we switch off the interactions, even if these do not have to be small. Every
sub-system then evolves independently. Imagine furthermore that some form of \emph{information
loss} takes place. Then, as was further motivated in Ref.~\cite{disdet}, we suspect that the
evolution in each domain will become \emph{periodic}.

Thus, we are led to consider the case where we have one or more independent, periodic variables
\(q_i(t)\). Only at a later stage, coupling between these variables will have to be introduced in
order to make them observable to the outside world. Thus, the introduction of periodic variables is
an essential ingredient of our theory, in addition to being just a useful exercise.

Consider a single periodic variable: \be {\pa q\over\pa t}=\w\ ,\eel{eomperiodic} while the state \(\{q=2\pi\}\) is
identified with the state \(\{q=0\}\). Because of this boundary condition, the associated operator \(p=-i\pa/\pa q\) is
quantized: \be p\iss 0,\,\pm1,\,\pm2,\,\cdots\ . \eel{quantump} The inessential additive coefficient \(g(q)\) of
Eq.~\eqn{hamil} here has to be real, because of Eq.~\eqn{hermitean}, and as such can only contribute to the
unobservable phase of the wave function, which is why we permit ourselves to omit it:\be H=\w\,p=\w\,n\ ;\qquad n\iss
0,\,\pm 1,\,\pm 2,\,\cdots\ . \eel{hamilint} If we would find a way to dispose of the negative energy states, this
would just be the Hamiltonian of the quantum harmonic oscillator with internal frequency \(\w\) (apart from an
inessential constant \(\half\w\)).\begin{quotation}
\noindent \textit{Theorem} \\
Consider any probability distribution \(W(q)\) that is not strictly vanishing for any value of
\(q\), that is, a strictly positive, real function of \(q\). Then a complex wave function \(\j(q)\)
can be found such that \be W(q)=\j^*(q)\j(q)\ , \eel{Wpsipsi} and \(\j(q)\) is a convergent linear
composition of eigenstates of \(H\) with non-negative eigenvalues only. \end{quotation}
\def\PP{{\cal P}} \noindent The proof is simple mathematics. Write \be \j(q)=\exp(\a(q)+i\b(q))\
,\qquad z=e^{iq}\ . \eel{psiz} Choose \(\a+i\b\) to be en entire function within the unit circle of
\(z\). Then an elementary exercise in contour integration yields, \be\fl\qquad
\a(q)=\half\log(W(q))\ ;\quad\b(q)=\b_0-\PP\oint{\dd q'\over 2\pi}\,{1+\ \cos(q'-q)\over
\sin(q'-q)}\,\a(q')\ , \eel{betaalpha} where \(\PP\) stands for the principal value, and \(\b_0\)
is a free common phase factor. In fact, Eq.~\eqn{betaalpha} is not the only function obeying our
theorem, because we can choose any number of zeros for \(\j(z)\) inside the unit circle and then
again match \eqn{Wpsipsi}. One concludes from this theorem that no generality in the function \(W\)
is lost by limiting ourselves to positive energy eigenfunctions only.

In fact, we may match the function \(W\) with a wave function \(\j\) that has a zero of an
arbitrary degree at the origin of \(z\) space. This way, one can show that the lowest energy state
can be postulated to be at any value of \(E\).

In this paper, however, we shall take a different approach. We keep the negative energy states, but
interpret them as representing the bra states \(\bra\j|\). These evolve with the opposite sign of
the energy, since \(\bra\j(t)|=e^{+iHt}\bra\j(0)|\). As long as we keep only one single periodic
variable, it does not matter much what we do here, since energy is absolutely conserved. The case
of two or more oscillators is more subtle, however, and this we consider in the next section.

In this bra-ket formalism, it will be more convenient to tune the energy of the lowest ket state at
\(\half\w\). The kets \(|n\ket\) and bras \(\bra n|\) have \(E_n=(n+\half)\w\). The time evolution
of the bras goes as if \(E_n=-(n+\half)\w\), so that we have a sequence of energy values ranging
from \(-\infty\) to \(\infty\).

\newsec{Two (or more) harmonic oscillators}\label{twoormore.sec}

As was explained at the beginning of Section \ref{harmosc.sec}, we expect that, when two periodic
variables interact, again periodic motion will result. This may seem to be odd. If the two periods,
\(\w_1\) and \(\w_2\) are incommensurate, an initial state will never exactly be reproduced. Well,
this was before we introduced information loss. In reality, periodicity will again result. We will
show how this happens, first by considering the quantum harmonic oscillators to which the system
should be equivalent, according to Section \ref{harmosc.sec}, and then by carefully interpreting
the result.

In Fig.~\ref{figure1.fig}, the states are listed for the two harmonic oscillators combined. Let
their frequencies be \(\w_1\) and \(\w_2\). The kets \(|n_1,\,n_2\ket=|n_1\ket|n_2\ket\) have
\(n_1\geq 0\) and \(n_2\geq 0\), so they occupy the quadrant labelled \(I\) in
Fig.~\ref{figure1.fig}. The bra states, in view of their time dependence, occupy the quadrant
labelled \(III\). The other two quadrants contain states with mixed positive and negative energies.
Those must be projected away. If we would keep those states, then any interaction between the two
oscillators would result in inadmissible mixed states, in disagreement with what we know of
ordinary quantum mechanics. So, although keeping the bra states is harmless because total energy is
conserved anyway, the mixed states must be removed. This is very important, because now we see that
the joint system cannot be regarded as a direct product. Some of the states that would be allowed
classically, must be postulated to disappear. We now ask what this means in terms of the two
periodic systems that we thought were underlying the two quantum harmonic oscillators.

\begin{figure}[ht] \setcounter{figure}{0}
\begin{quotation}
 \epsfxsize=80 mm\epsfbox{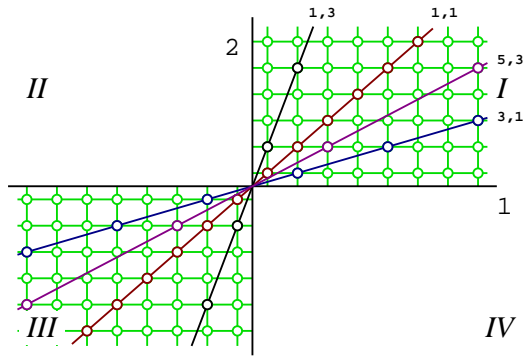}
  \caption{\footnotesize  Combining two harmonic oscillators. Tilted lines show sequences of
  spectral states again associated to harmonic oscillators. For further explanation, see text.}
  \label{figure1.fig}\end{quotation}
\end{figure} First, we wonder whether the spectrum of combined states will still be discrete. The classical, non interacting
system would only be periodic if the two frequencies have a rational ratio: \(p\,\w_1-q\,\w_2=0\), where \(p\) and \(q\) are
relative primes. The smallest period would be \(T=2\pi q/\w_1=2\pi p/w_2\), so that we would expect equally spaced energy levels
with spacings \(\w_1/q=\w_2/p\). Indeed, at high energies, we do get such spacings also in the quantum system, with increasing
degeneracies, but at lower energies many of these levels are missing. If the frequencies have an irrational ratio, the period of
the classical system is infinite, and so a continuous spectrum would have to be expected.

When two quantum harmonic oscillators are considered together, this does not happen. The spectrum is always discrete. In
Fig.~\ref{figure1.fig}, it is indicated how to avoid having missing states and variable degeneracies. We see that actually full
series of equally spaced energy levels still exist:\begin{quotation}\noindent \emph{At any given choice of a pair of odd relative
primes p and q, we have a unique series of bra- and ket states with energies} \(\w_{pq}(n+\half)\), with
\(\w_{pq}=p\,\w_1+q\,\w_2\).\end{quotation} It is easy to see that these sequences are not degenerate, that all odd relative
prime pairs of integers \((p,\,q)\) occur exactly once, and that all states are represented this way: \be
E_{n_1,\,n_2}=(n_1+\half)\,\w_1+(n_2+\half)\,\w_2=(n+\half)(p\,\w_1+q\,\w_2)\ ;\crl{frequences} {2n_1+1\over 2n_2+1}={p\over q}\
. \eel{ratios} Some of these series are shown in the Figure.

We see that, in order to reproduce the quantum mechanical features, that is, to avoid the unphysical states where one energy is
positive and the other negative, we have to combine two periodic systems in such a way that a new set of periodic systems arises,
with frequencies \(\w_{pq}\). Only then can one safely introduce interactions of some form. Conservation of total energy ensures
that the bra and ket states cannot mix. States where one quantum oscillator would have positive energy and one has negative
energy, have been projected out.

\begin{figure}[ht]
\begin{quotation}
 \epsfxsize=60 mm\epsfbox{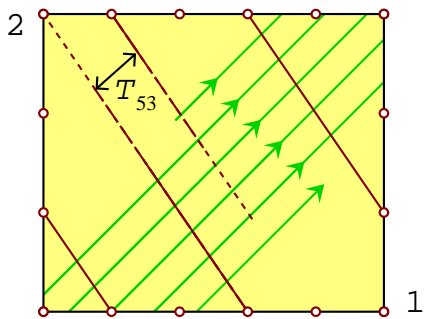}
  \caption{\footnotesize  The equivalence class for oscillators in the case \((p,\,q)=(5,\,3)\).
  Lines with arrows pointing right and up: time trajectories of individual points. Solid and broken
  lines going downwards: (part of) the (5,\,3) equivalence class at \(t=0\). For further explanation, see text.}
  \label{figure2.fig} \end{quotation}
\end{figure}

But how can such a rearrangement of the frequencies come about in a pair of classical periodic systems? Indeed, why are these
frequencies so large, and why are they labelled by odd relative primes? In Fig.~\ref{figure2.fig} the periodicities are displayed
in configuration space, \(\{q_1,\,q_2\}\). The combined system evolves as indicated by the arrows. The evolution might not be
periodic at all. Consider now the \((5,\,3)\) mode. We can explain its short period \(T_{53}=2\pi/\w_{53}\) only by assuming that
the points form equivalence classes, such that different points within one equivalence class are regarded as forming the same
`quantum' state. If all points on the lines shown in Fig.~\ref{figure2.fig} (the ones slanting downwards) form one equivalence
class, then this class evolves with exactly the period of the oscillator whose frequency is \(\w_{53}\).

In principle, this equivalence class can be formed in one of two ways: the information concerning the location of a point on this
line is lost, either because there is an inherent information loss mechanism, implying that two different states may actually
evolve to become the same state, or it could simply be that this information cannot be transmitted to macroscopically observable
quantities. One could imagine a renormalization group technique that relates microscopic states to states at much larger distance
scales. Not all data are being faithfully transmitted in the procedure. This latter option will later be dismissed as being
impractical; it is highly revealing to assume explicit information loss.

For the time being, imagine that information loss takes place by means of processes that are random, uncontrollable or impossible
to follow in detail, that cause our data point to fluctuate along the line of its equivalence class. The line itself moves with
the deterministic speed of the original oscillators.

Observe in Figure~\ref{figure2.fig} that, in case \((p,\,q)=(5,\,3)\), due to these fluctuations, five points of system 1 alone
now form a single equivalence class, and three points in system 2. This is because we have assigned 5 quanta of energy to system
1 for every three quanta of energy of system 2. More generally, we could represent this situation with the wave function \be
\j_{pq}=e^{i\,(n+{1\over 2})\,\left(p\,q_1+q\,q_2-\w_{pq}t\right)}\,e^{-{1\over 2}\,i( q_1+ q_2 )}\ ,\eel{psipq} where both
variables \(q_{1,2}\) were taken to be periodic with periods \(2\pi\). The \((p,\,q)\) equivalence classes appear to be defined
by the condition \be p\,q_1+q\,q_2\iss\hbox{Constant}\,,\eel{equclass} and this means that the \(n\)-dependent part of
 the wave function \eqn{psipq} has a the same phase all over the entire equivalence class, if we may assume that the second term in
 Eq.~\eqn{psipq}, arising from the vacuum fluctuations \(\half\,\w\,t\), may be ignored.

To describe the equivalence classes it is helpful to introduce time variables \(t_a\) for the subsystems \(a=1,2,\,\cdots\) in
terms of their \emph{unperturbed} evolution law, \(q_a=\w_at_a\). Then, writing \(E_1=p\,\w_1\,,\ E_2=q\,\w_2\), one can
characterize the equivalence classes as \be E_1\,\d\,t_1+E_2\,\d\,t_2\iss 0\ ,\eel{equivclasst} which means that the reactions
that induce information loss cause \(q_a\) to speed up or slow down by an amount \(\pm\d\,t_a\) obeying this equation. One can
easily generalize this result for many coexisting oscillators. They must form equivalence classes such that fluctuating time
differences occur that are only constrained by \be \sum_aE_a\,\d\,t_a=0\ ,\eel{multiclass} which also are the collections of
points that have the same phase in their quantum wave functions. We conclude that, in the ontological basis \(\{|\vec
q\,\ket\}\), \emph{all states \(|\vec q\,\ket\) which have the same phase in the wave function \(\bra\vec q|\j\ket\)} (apart from
a fixed, time independent term), \emph{form one complete equivalence class}.

\newsec{Energy and Hamiltonian}\label{hamilton.sec} In the previous section, it was derived that the energies of the various
oscillators determine the shape of the equivalence classes that are being formed. However, this would require energy to be a
beable, as defined in Section~\ref{beables.sec}. Of course, the Hamiltonian, being the generator of time evolution, cannot be a
beable. It is important to notice here, that the parameters \(p\) and \(q\) defining the equivalence classes as in
Section~\ref{twoormore.sec}, are not exactly the energies of \(q_1\) and \(q_2\); the Hamiltonian eigenvalues are \be
H_1=(n+\half)p\,\w_1\ ;\qquad H_2=(n+\half)q\,\w_2\ ,\eel{energiespq} with a common integral multiplication factor \(n+\half\).
This \(n\) indeed defines the Hamiltonian of the orbit of the equivalence class. Generalizing this, the relation between the
energies \(E_i\) in Eqs.~\eqn{equivclasst} and \eqn{multiclass} and the Hamiltonian \(H\), is \be H=(n+\half)E\ ,\eel{HnE} where
\(n\) defines the evolution of a single clock that monitors the evolution of the entire universe.

Now that the relative primes \(p\) and \(q\) have become beables, we may allow for the fact that
the periods of \(q_1\) and \(q_2\) depend on \(p\) and \(q\) as a consequence of some non-trivial
interaction. But there is more. We read off from Fig.~\ref{figure2.fig}, that \(p\) points on the
orbit of \(q_1\) in fact belong to the same equivalence class. Assuming that the systems \(1\) and
\(2\) that we started off with, had been obtained again by composing other systems, we must
identify these points. But this forces us to redefine the original periods by dividing these by
\(p\) and \(q\), respectively, and then we end up with two redefined periodic systems that are
combined in the one and only allowed way: \be p=q=1\ . \eel{pqone} Only a single line in
Fig.~\ref{figure1.fig} survives: the diagonal.

The picture that emerges is the following. We are considering a collection of variables \(q_a\), each being periodic
with different periods \(T_a=2\pi/\w_a\). They each are associated with a positive beable \(E_a\), such that
\(E_a=\w_a\). The interactions will be such that the total energy \(E=\sum_aE_a\) is conserved. Now as soon as these
variables are observed together (even if they do not interact), an uncontrollable mixing mechanism takes place in such
a way that the variables are sped up or slowed down by time steps \(\d t_a\) obeying Eq.~\eqn{multiclass}, so that, at
any time \(t\), all states obeying \be \sum_aE_at_a=\Big(\sum_aE_a\Big)t\ , \eel{totequclass} form one single
equivalence class.

The evolution and the mixing mechanism described here are entirely classical, yet we claim that
such a system turns into an acceptable quantum mechanical theory when handled probabilistically.
However, we have not yet introduced interactions.

\newsec{Interactions}\label{interaction.sec} We are now in a position to formulate the problem of
interacting systems. Consider two systems, labelled by an index \(a=1,\,2\). System \(a\) is
characterized by a variable \(q_a\in[\,0,\,2\pi)\) and a discrete index \(i=1,\,\cdots,\,N_a\),
which is a label for the spectrum of states the system can be in. Without the interaction, \(i\)
stays constant. Whether the interaction will change this, remains to be seen.

The frequencies are characterized by the values \(E_a^i=\w_a^i\), so that the periods are \(T_a^i=2\pi/\w_a^i\).
Originally, as in Section \ref{twoormore.sec}, we had \(\w_a^i=p_a\w_a\), where \(p_1=p\) and \(p_2=q\) were relative
primes (and both odd), but the periods \(\w_a\) are allowed to depend on \(p_a\), so it makes more sense to choose a
general spectrum to start with.

The non-interacting parts of the Hamiltonians of the two systems, responsible for the evolution of each, are described
by \be H_a^0|n_a,\,i\ket=(n_a+\half)E_a^i|n_a,\,i\ket\ ,\eel{Hama} where the integer \(n_a=-\infty,\,\cdots,\,+\infty\)
is the changeable generating the motion along the circle with angular velocity \(\w_a\). We have \be n_a=-i\pa/\pa q_a\
. \eel{nddq} \def\tot{\mathrm{tot}}

The total Hamiltonian describing the evolution of the combined, unperturbed, system is not \(H^0_1+H^0_2\), but \be
H_\tot^0=(n_\tot+\half)(E_1^i+E_2^j)\ ,\eel{htotnul} where \(i,\,j\) characterize the states 1 and 2, but we have a
single periodic variable \(q_\tot\in[\,0,\,2\pi)\), and \be n_\tot=-i{\pa/\pa q_\tot}\ . \eel{ntot} In view of
Eq.~\eqn{equclass}, which here holds for \(p=q=1\), we can define \be q_\tot= q_1+q_2\ ,\eel{qtot} while \(q_1-q_2\)
has become invisible. We can also say,\be n_1=n_2=n_\tot\ . \eel{nisnn} \def\intt{\mathrm{int}}

An interacting system is expected to have perturbed energy levels, so that its Hamiltonian should become \be H^0+H^\intt
=(n_\tot+\half)(E_1^i+E_2^j+\d E^{ij})\ ,\eel{hinteract} where \(\d E^{ij}\) are correction terms depending on both \(i\) and
\(j\). This is realized simply by demanding that the beables \(E_1^i\) and/or \(E_2^j\) get their correction terms straight from
the other system. This is an existence proof for interactions in this framework, but, at first sight, it appears not to be very
elegant. It means that the velocity \(\w_1^{ij}\) of one variable \(q_1\) depends on the state \(j\) that the other variable is
in, but no matrix diagonalization is required. Indeed, we still have no transitions between the different energy states \(i\). It
may seem that we have to search for a more general interaction scheme. Instead, the scheme to be discussed next differs from the
one described in this section by the fact that the energies \(E\) cannot be read off directly from the state a system is in, even
though they are beables. The indices \(i,\,j\) are locally unobservable, and this is why we usually work with superimposed
states.

\newsec{Limit cycles}\label{cycles.sec} Consider the evolution following from a given initial configuration at \(t=t_0\),
having an energy \(E\). Let us denote the state at time \(t\) by \(F(t)\). The equivalence classes \(|\j(t)\ket\)
defined in Section~\ref{twoormore.sec} are such that the state \(F(t)\) is equivalent to the state \(F(t+\D t)\), where
\be \D t=h/E\ , \eel{Deltat} in which \(h\) is Planck's constant, or: \be|\j(t)\ket\deff\{F(t+n\D t)\,,\ \forall
n|_{n\ge n_1}\}\ ,\eel{defpsi} for some \(n_1\). Let us now assume that the equivalence indeed is determined by
information loss. That the states in Eq.~\eqn{defpsi} are equivalent then means that there is a smallest time \(t_1\)
such that \be F(t_1+n\D t)=F(t_1)\ ,\ \forall n\ge0\ . \eel{smallestt} Thus, the system ends in a \emph{limit cycle}
with period \(\D t\).

\begin{figure}[ht]
\begin{quotation}
 \epsfxsize=45 mm\epsfbox{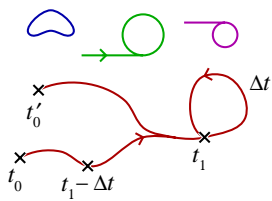}
  \caption{\footnotesize  Configuration space showing the limit cycles of an evolving system,
  indicating the times \(t_0\), \(t_1\) and the period \(\D t\) of a limit cycle. The points at
  \(t_0\) and \(t'_0\) form one equivalence class.}
  \label{figure3.fig} \end{quotation}
\end{figure}

One now may turn this observation around. A closed system that can only be in a finite number of different states, making
transitions at discrete time intervals, would necessarily evolve back into itself after a certain amount of time, thus exhibiting
what is called a Poincar\'e cycle. If there were no information loss, these Poincar\'e cycles would tend to become very long,
with a periodicity that would increase exponentially with the size of the system. If there is information loss, for instance in
the form of some dissipation effect, a system may eventually end up in Poincar\'e cycles with much shorter periodicities. Indeed,
time does not have to be discrete in that case, and the physical variables may form a continuum; there could be a finite set of
stable orbits such that, regardless the initial configuration, any orbit is attracted towards one of these stable orbits; they
are the limit cycles. The energy of a state is then simply defined to be given by Eq.~\eqn{Deltat}, or \(E\deff h/P\), where
\(P=\D t\) is the period of the limit cycle, and \(h\) is Planck's constant.

Since this period coincides with the period of the wave function, we now deduce a physical interpretation of the phase of
the wave function: \emph{The phase of a wave function} (in the frame of energy eigenstates) \emph{indicates where in the
limit cycle the state will be.}

In general, we will have a superposition of many possibilities, and so we add to this the interpretation of the amplitude of
the wave function: the absolute value of the amplitude in the frame of energy eigenstates indicates the probability that a
particular limit cycle will be reached. Thus, we have reached the exact physical meaning of a quantum wave function.

We identified any deterministic system having information loss, with a quantum mechanical system evolving with a Hamiltonian
defined by Eq.~\eqn{HnE}. However, in order to obtain a realistic model, one has to search for a system where the energy is
extrinsic. With this, we mean that the universe consists of subsystems that are weakly coupled. Uncoupled systems are
described as in Section~\ref{twoormore.sec}; weakly coupled systems must be such that the limit cycles of the combination
(12) of two systems (1) and (2) must have periods \(P_{12}\) obeying \be {1\over P_{12}}\approx{1\over P_1}+{1\over P_2}\
,\eel{combn} which approaches the exact identity in the limit of large systems being weakly coupled. This is the energy
conservation law.

\newsec{Discussion}\label{disc.sec}
When we attempt to regard quantum mechanics as a deterministic system, we have to face the problem of the positivity of
the Hamiltonian, as was concluded earlier in Refs~\cite{disdet}\cite{Elze1}\cite{Kleinertetal}. There, also, the
suspicion was raised that information loss is essential for the resolution of this problem. In this paper, the
mathematical procedures have been worked out further, and we note that the deterministic models that we seek must have
short limit cycles, obeying Eq.~\eqn{Deltat}. Short limit cycles can easily be obtained in cellular automaton models
with information loss, but the problem is to establish the addition rule \eqn{combn}, which suggests the large
equivalence classes defined by Eq.~\eqn{equclass}. We think that the observations made in this paper are an important
step towards the demystification of quantum mechanics.

We found that the energy eigenstates of a quantum system correspond to the limit cycles of the deterministic model. If
\(P\) is the period of the limit cycle, then the energy \(E\) of this state is \(E=h/P\) (see Eq.~\ref{Deltat}).

\def\OO{{\cal O}}\def\NN{{\cal N}}\def\PP{{\cal P}}\def\EE{{\cal E}}

In models with more or less random evolution laws, one can guess the distribution of the periods of the limit cycles.
In the Appendix, we derive the distribution of limit cycles with periods \(\D t=P\) for a ``completely random" model,
which we define to be a model where the mapping \(F(t)\ra F(t+1)\) is chosen completely randomly for every \(F(t)\),
independently of how many other states \(F'(t)\) might map into the same state \(F(t+1)\). It is found that the
distribution of the cycles may then be expected to be logarithmic, which leads to a logarithmic energy spectrum: the
energy eigenstates are a Poisson distribution on a logarithmic scale: \be \r(E)\dd E=\dd E/E=\dd\log E \ ,
\eel{energydistr} with cutoffs at \be E_\mathrm{min}=\OO(h/\sqrt{2\NN})\ ,\qquad E_\mathrm{max}=\OO(1/\d t)\ ,
\eel{Eminmax} where \(\NN=\OO(e^V)\) is the total number of possible states and \(\d t\) is some cutoff in time that
our system may have. This is not a realistic energy spectrum for a quantum field theory, so we must conclude that
realistic models will have to be far from random.

Cellular automaton models can be written down that show a rapid convergence towards small limit cycles, starting from
any state \(F(0)\). Conway's ``game of life"\cite{Conway} is an example, although that also features `glider
solutions', which are structures that are periodic, but they move forward when released in an empty region, so that
they are not limit cycles in the strict sense. It must be emphasized, however, that Conway's game of life will not
serve as a model generating quantum mechanics. In a model generating quantum mechanics, the vacuum state is the state
with the \emph{longest} limit cycle, since it has the lowest energy. Thus, the empty state in Conway's game of life
would carry more energy than its glider solutions.

The limit cycles of a random model are too long, those of cellular automata such as Conway's Game of Life are too short. In the
real universe we have a small number of massless particle species, and many more massive ones, generating a rich spectrum of
energies all very low compared to the Planck energy. If we assume that the Planck time would be Nature's natural time scale, then
we observe that there must be many limit cycles whose periods are very long compared to the Planck time. Our universe appears to
be built in such a way that, as soon as several of these cyclic limiting solutions are allowed to interact, new limit cycles will
be reached with shorter periods, due to information loss.

In cellular automaton models, one might be able to mimic such a situation best by introducing a nearly conserved, positive
quantity resembling energy, which can be seen statistically to decrease slowly on the average, so that the most chaotic initial
states relax into more organized states that can easily end up in a limit cycle. The more chaotic the initial state, the smaller
the period of its eventual limit cycle is expected to be, but there are many special initial states with very long limit cycles:
the low energy states.

States of interest, with which we might attempt to describe the universe as we know it, must be very far away from any limit
cycle. They are also far away from the strictly stationary eigenstates of the Hamiltonian. This means that we \emph{do not yet
know} which of the numerous possible limit cycles our universe will land into. This is why we normally use wave functions that
have a distribution of amplitudes in the basis of the Hamiltonian eigenmodes. The squares of these amplitudes indicate the
probability that any particular limit cycle will be reached. Also note that, according to General Relativity, taking into account
the negative energies in the gravitational potentials, the total energy of the universe should vanish, which means that the
entire universe might never settle for any limit cycle, as is indeed suggested by what we know of cosmology today: the universe
continues to expand. The limit cycles mentioned in this paper refer to idealized situations where small sections of the universe
are isolated from the rest, so as to be able to define their energies exactly. Only when a small part of the universe is sealed
off from the rest, it is destined to end up in a limit cycle.

It may be of importance to note that our definition of energy, as being the inverse of the period of the limit cycle, supplies us
with an absolute scale of energy: it is not allowed to add a constant to it. Moreover, the energy \(E\) in Eq.~\eqn{HnE}, as
opposed to the Hamiltonian \(H\), is a beable. Thus it is allowed to couple it to gravity by imposing Einstein's equations. If
indeed the vacuum has a limit cycle with a large period, it carries a very low energy, and this is why we suspect that the true
resolution of the cosmological constant problem\cite{Nobb} will come from deterministic quantum mechanics rather than some
symmetry principle\cite{NobbGtH}. Earlier, the cosmological constant has been considered in connection with deterministic quantum
mechanics by Elze\cite{Elze2}. The fact that the observed cosmological constant appears to be non-vanishing implies that a finite
volume \(V\) of space will have a largest limit cycle with period \be P={8\pi hG\over\L V}\ , \eel{vacperiod} which is of the
order of a microsecond for a volume of a cubic micron. If \(\L\) were negative we would have had to assume that gravity does not
exactly couple to energy.

Lorentz transformations and general coordinate transformations have not been considered in this paper. Before doing
that, we must find models in which the Hamiltonian is indeed extensive, that is, it can be described as the integral of
an Hamiltonian density \(T_{00}(\vec x,t)\) over 3-space, as soon as the integration volume element \(\dd^3\vec x\) is
taken to be large compared to the `Planck volume'. When that is achieved, we will be only one step away from generating
locally deterministic quantum field theories.

What can be said from what we know presently, is that a particle with 4-momentum \(p_\m\) must represent an equivalence class
that contains all translations \(x^\m\ra x^\m+\D x^\m\) with \(p_\m\,\D x^\m = n\,h\), where \(n\) is an integer. Note that a
limit cycle having this large transformation group as an invariance group is hard to imagine, which probably implies that this
particular limit cycle will take an infinite time to be established. Indeed, a particle in a fixed momentum state occupies the
entire universe, and we already observed that the entire universe will never reach a limit cycle.

In the real world, we have only identified the observable quantum states, which we now identify with the equivalence classes
of ontological states. We note that the physical states of the Standard Model in fact are also known to be gauge equivalence
classes, Local gauge transformations modify our description of the dynamical variables, but not the physical observations.
It is tempting to speculate that these gauge equivalence classes (partly) coincide with the equivalence discussed in this
paper, although our equivalence classes are probably a lot larger, which may mean that many more local gauge symmetries are
still to be expected.

It is even more tempting to include here the gauge equivalence classes of General Relativity: perhaps local coordinate
transformations are among the dissipative transitions. In this case, the underlying deterministic theory might not be
invariant under local coordinate transformations, and here also one may find novel approaches towards the cosmological
constant problem and the apparent flatness of our universe.

Our reason for mentioning virtual black holes being sources of information loss might require further explanation. Indeed, the
\emph{quantum mechanical} description of a black hole is not expected to require information loss (in the form of quantum
decoherence); it is the corresponding classical black hole that we might expect to play a role in the ontological theory, and
that is where information loss is to be expected, since classical black holes do not emit Hawking radiation. As soon as we turn
to the quantum mechanical description in accordance to the theory explained in this paper, a conventional, fully coherent quantum
description of the black hole is expected.

Although we do feel that this paper is bringing forward an important new approach towards the interpretation of Quantum
Mechanics, there are many questions that have not yet been answered. One urgent question is how to construct explicit models in
which energy can be seen as extrinsic, that is, an integral of an energy density over space. A related problem is how to
introduce weak interactions between two nearly independent systems. Next, one would like to gain more understanding of the
phenomenon of (destructive) interference, a feature typical for Quantum Mechanics while absent in other statistical theories.

\appendix

\newsec{The random deterministic model}\label{app.sec} For simplicity, take the duration \(\d t\)
of one time step, in Eq.~\eqn{clock}, to be 1. Consider a completely random mapping \(F(t)\ra F(t+1)\). This means that,
using a random number generator, some rule has been established to define \(F(t+1)\) for any given \(F(t)\). The rule does
not depend on \(t\). In general, the mapping will not at all be one-to-one. Let the total number of states \(F(t)\) be
\(\NN\). In general, \(\NN\) wil grow exponentially with the volume \(V\) of the system. We claim that the distribution of
limit cycles may be expected to be as described in Section \ref{disc.sec}. The argument goes as follows.

At \(t=0\), take an arbitrary element of the space of states, \(F(0)\). The series \(F(1),\,F(2),\,\cdots\) will end up in a
limit cycle, which means that there is a time \(T\) that is the smallest time with the property that \(F(T)=F(t_1)\) for
some \(0\le t_1< T\). The length of the limit cycle is \(P=T-t_1\), see Fig.~\ref{figure3.fig}. We consider the case that
\(\NN\gg 1\).

Let \(Q(x)\) be the probability that \(T>x-1\), given the state \(F(0)\). The probability that \(T=x\) is then equal to
\((x/\NN)Q(x)\), so that \be\fl\qquad {\dd \over\dd x}Q(x)=-{x\over\NN}\,Q(x)\ ,\qquad Q(0)=1\qquad\ra\qquad Q(x)=e^{-x^2/
(2\NN)}\ . \eel{Qx} In general, \(t_1>0\), so that the state \(F(0)\) itself is not a member of the limit cycle. Apparently the
limit cycles contain only a small subset of all states. The probability that the element \(F(0)\) actually happens to sit in the
limit cycle with length \(P\) (to be called \(C(P)\)), is the probability that \(t_1=0\), or \be \PP\Big(F(0)\in
C(P)\Big)={1\over\NN}\,Q(P)\ . \eel{FinC} This is understood as follows: \(Q(P)\) is the chance that the cycle, starting from
\(F(0)\) did not close earlier, and \(1/\NN\) is the probability that the \(P^\mathrm{th}\) state happens to coincide with
\(F(0)\).

Since \(C(P)\) contains \(P\) states, and the total number of states is \(\NN\), one derives that the expectation value of
the number of limit cycles with length \(P\) is \be \EE(P)={1\over P}\,e^{-P^2/(2\NN)}\ . \eel{expP} From this, we derive
the distribution of periods \(P\) to be \be \r(P)\dd P={\dd P\over P}\,e^{-P^2/(2\NN)}\ . \eel{distrP} For all periods \(P\)
that are small compared to \(\sqrt\NN\), the exponent can be ignored, and since \(E=h/P\), Eq.~\eqn{energydistr} follows.
The largest period is of order \(\sqrt\NN\), and the smallest one is the fundamental time unit, which here was taken to be
one.

\end{document}